\documentclass[prl, aps, superscriptaddress, twocolumn, amsmath, amssymb, 10pt]{revtex4-2}

\usepackage{amsmath,graphicx,color}
\usepackage{bm}
\usepackage{natbib}
\usepackage{placeins}
\setcitestyle{square}
\usepackage{lipsum}
\usepackage{braket}
\usepackage{leftindex}
\usepackage{makerobust}
\usepackage{xcolor}
\usepackage{hyperref}
\hypersetup{%
 colorlinks,
 breaklinks=true,
 plainpages=false,%
 citecolor=blue,
 linkcolor=blue,
 urlcolor=blue,
 bookmarksopen=true,%
 bookmarksnumbered=false,%
 bookmarksdepth=5%
}

\newcommand{\FSTx}{FeSe$_{1-x}$Te$_{x}$}

\newcommand{\makeauthor}[2]{\newcommand{#1}[1]{{
 \sffamily\color{#2}{%
 \bfseries\begingroup\escapechar=-1\edef\x{\endgroup\string#1}\x:
 } ##1}}
 \MakeRobustCommand#1}
\makeauthor{\dkm}{red}
\makeauthor{\rew}{blue}
\makeauthor{\jb}{purple}

\begin{document}

\title{Majorana Zero Modes in a Heterogenous Structure of Topological and Trivial Domains in FeSe$_{1-x}$Te$_x$} 

\author{Prashant Gupta}
\author{Jasmin Bedow}
\affiliation{Department of Physics, University of Illinois Chicago, Chicago, IL 60607, USA}
\author{Eric Mascot}
\affiliation{School of Physics, University of Melbourne, Parkville, VIC 3010, Australia}
\author{Dirk K. Morr}
\affiliation{Department of Physics, University of Illinois Chicago, Chicago, IL 60607, USA}
\date{\today}

\begin{abstract}
We propose that the existence of vortices in FeSe$_{1-x}$Te$_x$ with and without Majoarana zero modes (MZMs) can be explained by a heterogeneous mixture of strong topological and trivial superconducting domains, with only vortices in the former exhibiting MZMs.  We identify the spectroscopic signatures of topological and trivial vortices and show that they are necessarily separated by a domain wall harboring Majorana edge modes. We demonstrate that when a vortex is moved from a trivial to a topological domain in real time, a domain wall Majorana edge mode is transferred to the vortex as an MZM. 
\end{abstract}


\maketitle

{\bf Introduction.~} 
Over the last few years, strong evidence for the existence of topological surface superconductivity in the iron-chalcogenide superconductor \FSTx has emerged, ranging from the observation of a surface Dirac cone \cite{Zhang2018,Rameau2019,Zaki2019,Yangmu2021}, to that of zero-energy states, proposed to be Majorana zero modes (MZMs), in vortex cores \cite{Wang2018,Machida2019,Kong2019,Zhu2020} and of putative Majorana edge modes (MEMs) at domain walls \cite{Wang2020}. The microscopic origin of these observations has been attributed to topological surface superconductivity arising (a) from a Fu-Kane like mechanism \cite{Fu2008,Wang2015,Wu2016,Xu2016,Zhang2018} of proximity induced superconductivity in the surface Dirac cone of a three-dimensional topological insulator, or (b) from a combination of experimentally observed surface magnetism \cite{Rameau2019,Zaki2019,Yangmu2021,McLaughlin2021}, a Rashba spin-orbit interaction, and a hard $s_{\pm}$-wave superconducting gap \cite{Mascot2022,Wong2022,Xu2023}.  One of the most puzzling, and yet unexplained, observations has been that the presence of zero-energy states is not ubiquitous in all vortices \cite{Wang2018,Machida2019,Kong2019,Zhu2020}, which has been attributed to the presence of disorder and formation of domains \cite{Wu2021,Barik2022} and/or hybridization effects \cite{Chiu2020}. However, the observation of surface \cite{Rameau2019,Zaki2019,Yangmu2021,McLaughlin2021} as well as bulk ferromagnetism \cite{Roppongi2025} in \FSTx, which can destroy a Fu-Kane type topological superconductor \cite{Wu2021,Xu2023}, has renewed the interest in the nature of such domains, and their spectroscopic fingerprints.
\begin{figure}[!htbp]
    \centering
    \includegraphics[width=\columnwidth]{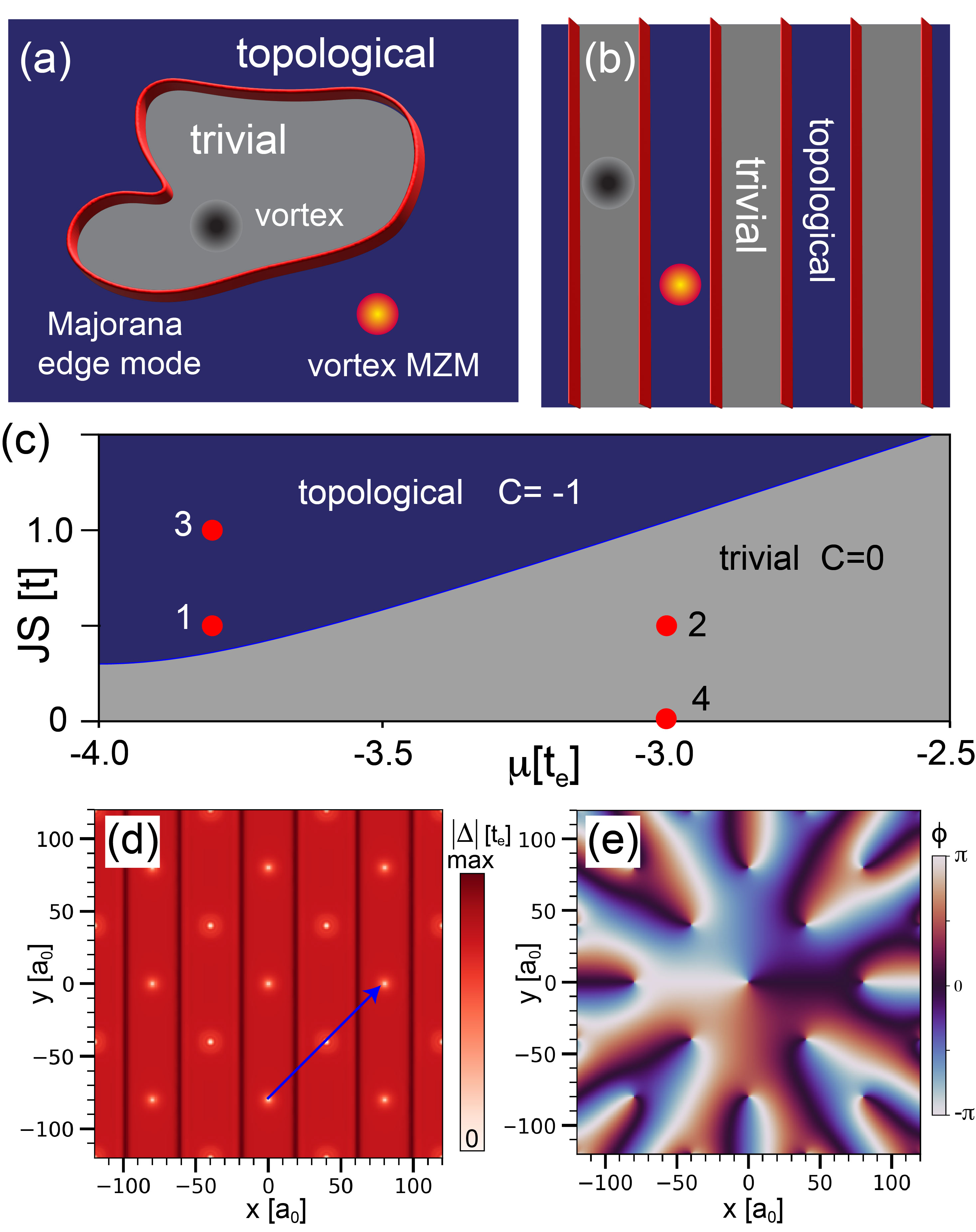}
    \caption{Schematic of the topological and trivial regions in (a) FeSeTe, and (b) the system considered here. (c) Topological phase diagram. Spatial dependence of the (d) magnitude and (e) phase of the superconducting order parameter, computed self-consistently. Parameters in (d) for the topological and trivial domain are given by points 1 and 2 in (c) with $\alpha = 0.2 t_e$, $W=4a_0$. These values of $\alpha$, $W$ are also used in Figs.~2,3.
    }
    \label{fig:Fig1}
\end{figure}

In this article, we study the spectroscopic signatures of a heterogeneous mixture of topological and trivial domains in FeSe$_{1-x}$Te$_x$ [see Fig.~\ref{fig:Fig1}(a)]. Motivated by the strong evidence for surface \cite{Rameau2019,Zaki2019,Yangmu2021,McLaughlin2021} and bulk ferromagnetism \cite{Roppongi2025} in \FSTx, we employ a scenario in which topological superconductivity arises from the interplay of ferromagnetism, a Rashba spin-orbit interaction, and a hard $s$-wave superconducting gap \cite{Mascot2022,Wong2022,Xu2023}. We identify the spectroscopic signatures of topological and trivial domains, and show that vortices in the former (topological vortices) host an MZM, while those in the latter (trivial vortices) do not. Since topological and trivial domains are separated by a domain wall harboring a Majorana edge mode, we predict that a $\mathrm{d}I/\mathrm{d}V$ line cut measured between a topological and trivial vortex will also reveal the presence of MEMs. The nature of this heterogeneous domain structure is qualitatively different from previous proposals \cite{Wu2021,Barik2022}, leading to distinct experimental differences. Finally, using a non-equilibrium formalism \cite{Bedow2022}, we demonstrate that when a vortex is moved from a trivial to a topological domain, a MEM is transferred from the domain wall to the vortex as a localized MZM. Our study thus predicts the detailed spectroscopic fingerprints that would allow future experiments to ascertain the existence of topological and trivial domains in \FSTx.

{\bf Theoretical Methods.~} To study a 2D superconductor containing both topological and trivial domains in the presence of magnetic vortices, we consider a previously employed model for \FSTx \cite{Mascot2022,Wong2022,Xu2023}, in which topological superconductivity arises from the interplay of $s$-wave superconductivity, Rashba spin-orbit coupling and ferromagnetism, as described by the Hamiltonian
\begin{align}
\mathcal{H} =& \; \sum_{{\bf r}, {\bf r}', \beta, \gamma} e^{\mathrm{i} \theta({\bf r}, {\bf r}')} c^\dagger_{{\bf r}, \beta} \left[ -t_e \delta_{\beta,\gamma} + \mathrm{i} \alpha \left(\hat{e}_{{\bf r}' - {\bf r}} \times \boldsymbol{\sigma} \right)^z_{\beta, \gamma} \right] c_{{\bf r}', \gamma} \nonumber \\
&-  \sum_{{\bf r}, \beta} \mu_{\bf r} c^\dagger_{{\bf r}, \beta} c_{{\bf r}, \beta}
 + \sum_{{\bf r}}  |\Delta_{\bf r}| \left( e^{\mathrm{i} \phi({\bf r})} c^\dagger_{{\bf r}, \uparrow} c^\dagger_{{\bf r}, \downarrow} +h.c. \right)  \nonumber \\ 
&- {\sum_{{\bf r},\beta}} J_{\bf r}  c^\dagger_{{\bf r}, \beta}  \left[{\bf S}_{\bf r} \cdot {\bm \sigma}\right]_{\beta,\gamma} c_{{\bf r}, \gamma}  \; . \label{eq:ham}
\end{align}
Here, $c^\dagger_{{\bf r}, \beta}$ creates an electron with spin $\beta$ at site ${\bf r}$, $t_e$ denotes the nearest-neighbor hopping parameter on a 2D square lattice, $\alpha$ is the Rashba spin-orbit coupling between nearest-neighbor sites ${\bf r}$ and ${\bf r}'$, $\mu_{\bf r}$ is the local chemical potential, $|\Delta_{\bf r}|$ is the magnitude of the $s$-wave superconducting order parameter (SCOP) at site ${\bf r}$, and $J_{\bf r}$ is the magnetic exchange coupling between the magnetic adatom with spin ${\bf S}_{\bf r}$ of magnitude $S$ at site ${\bf r}$  and the conduction electrons.  A ferromagnetic out-of-plane alignment of the magnetic moments then yields the topological phase diagram shown in Fig.~\ref{fig:Fig1}(c) \cite{Li2016,Rachel2017}. 
Below, we model topological and trivial regions through spatial variations in the chemical potential $\mu_{\bf r}$ and/or of the magnetic exchange coupling $J_{\bf r}$. While the length over which these physical quantities spatially vary between domains defines a domain wall, $W$, we find that the qualitative nature of our results presented below remains unaffected by changes in $W$. Moreover, vortices are implemented through the Peierls phase 
$\theta({\bf r}, {\bf r}')= \frac{\pi}{\phi_0} \int_{\bf r}^{\bf r'} {\bf A} \cdot \mathrm{d}{\bf l}$ arising from the vector potential ${\bf A}$ (for details, see Appendix B). The resulting spatially dependent magnitude $|\Delta_{\bf r}|$ and phase $\phi({\bf r})$ of the superconducting order parameter are calculated self-consistently (for details, see Appendix A).

{\bf Results.~}
Rather than considering a random structure of topological and trivial domains, as schematically shown in Fig.~\ref{fig:Fig1}(a), we consider for computational reasons the ribbon geometry of domains shown in Fig.~\ref{fig:Fig1}(b), with periodic boundary conditions in the $x$- and $y$-directions. 
In Figs.~\ref{fig:Fig1}(d) and (e), we present the self-consistently computed magnitude and phase of the superconducting order parameter, respectively. As expected, we find that the order parameter is suppressed in the center of the vortex. In addition, we find that the SCOP is slightly enhanced at the domain walls between the trivial and topological domains.
\begin{figure}[!htbp]
    \centering
    \includegraphics[width=\columnwidth]{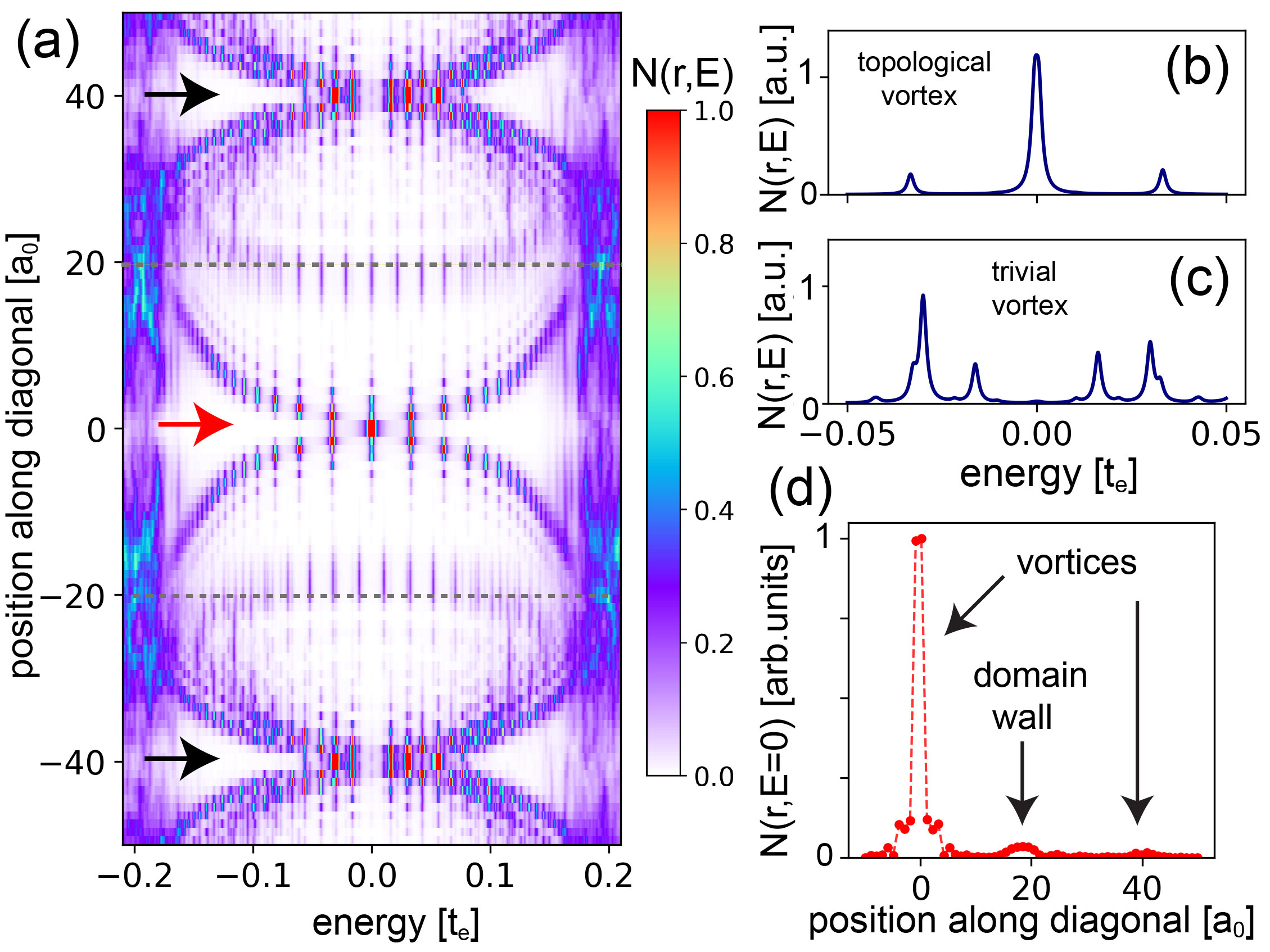}
    \caption{(a) Linecut of the LDOS between vortices in topological and trivial domains along the line indicated in Fig.~\ref{fig:Fig1}(d). The domain walls are indicated by gray dashed lines, while the position of the vortices in the topological (trivial) regions are indicated by red (black) arrows. Parameters for the topological and trivial domain are given by points 1 and 2 in Fig.~\ref{fig:Fig1}(c). Energy-dependent LDOS at the site of a (b) topological and (c) trivial vortex. (d) Line cut of the zero-energy LDOS between a topological and a trivial vortex.
    }
    \label{fig:Fig2}
\end{figure}
In Fig.~\ref{fig:Fig2}(a), we present the local density of states (LDOS) along a line-cut between vortices in topological and trivial domains [see blue arrow in Fig.~\ref{fig:Fig1}(d)]. As expected, we find that while the vortices in the topological domains host a zero-energy state (see red arrow), a Majorana zero mode, those in the trivial domain do not (see black arrows), as also evidenced by the energy-dependent LDOS at the site of a topological [Fig.~\ref{fig:Fig2}(b)] and trivial vortex [Fig.~\ref{fig:Fig2}(c)]. Moreover, since the Chern number changes at the domain wall (see gray dashed line) between the topological and trivial domains, the bulk boundary correspondence requires the existence of a Majorana edge mode. Due to the finite size of the supercell in our calculations, the MEM exhibits a finite discretization in energy \cite{Rachel2017}. Thus the proposal that the presence or absence of MZMs in vortex cores is related to vortices being located in topological or trivial domains necessarily requires that a Majorana edge mode be observable at the domain wall between these two regions. This is a necessary requirement for this proposal, which can be tested experimentally. In particular, a scanning tunneling microscopy (STM) experiment performing a zero bias linecut of the differential conductance, $\mathrm{d}I/\mathrm{d}V$ (which is proportional to the zero-energy LDOS) between a topological and a trivial vortex, as shown in Fig.~\ref{fig:Fig2}(d) would show not only a peak at the position of the topological vortex, and the absence thereof at the position of the trivial vortex, but would also show a peak related to the Majorana edge modes when the domain wall is crossed. The presence of the latter would therefore strongly support the conjecture, that the 
presence and absence of vortex core MZMs is related to the existence of topological and trivial domains. 
\begin{figure}[t]
    \centering
    \includegraphics[width=\columnwidth]{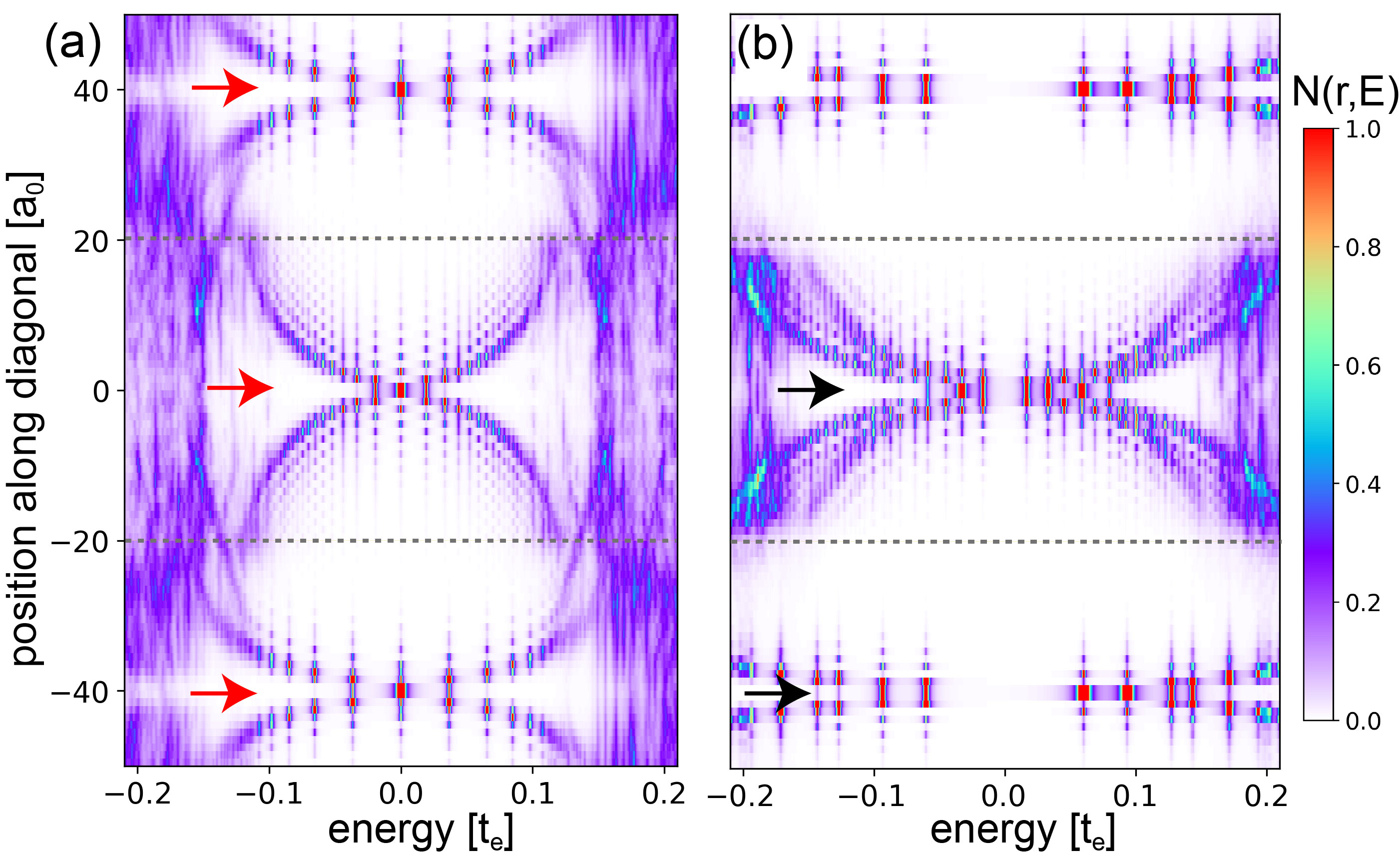}
    \caption{Linecut of the LDOS across (a) three topological and (b) three trivial domains along the line indicated in Fig.~\ref{fig:Fig1}(d). The domain walls are indicated by gray dashed lines, while the position of the vortices in the topological (trivial) regions are indicated by red (black) arrows. Parameters for the topological domains and trivial domains are given by points 1,3 and 2,4 in Fig.~\ref{fig:Fig1}(c), respectively. }
    \label{fig:Fig3}
\end{figure}
To further support this conclusion, we consider the case when variations in the chemical potential or the strength of magnetic moments modify the electronic structure of the domains, but not their topological (or trivial) nature. Consider, for example, the case when domains are formed because of variations in the strength of the magnetic moment $S$, but are still in the same topological phase. In this case, as follows from the linecut of the LDOS between vortices in different topological domains shown in Fig.~\ref{fig:Fig3}(a), all of the vortices possess an MZM at their center, and there exists no Majorana edge mode at the domain wall [see Fig.~\ref{fig:Fig3}(a)]. Similarly, when both types of domains are trivial, the vortices possess no MZMs, and no Majorana edge mode exists at the domain wall [see Fig.~\ref{fig:Fig3}(b)]. Thus, the simultaneous observation of a vortex core MZM and a Majorana edge mode can only be explained by the presence of topological and trivial domains.
 
\begin{figure}[!htbp]
    \centering
    \includegraphics[width=\columnwidth]{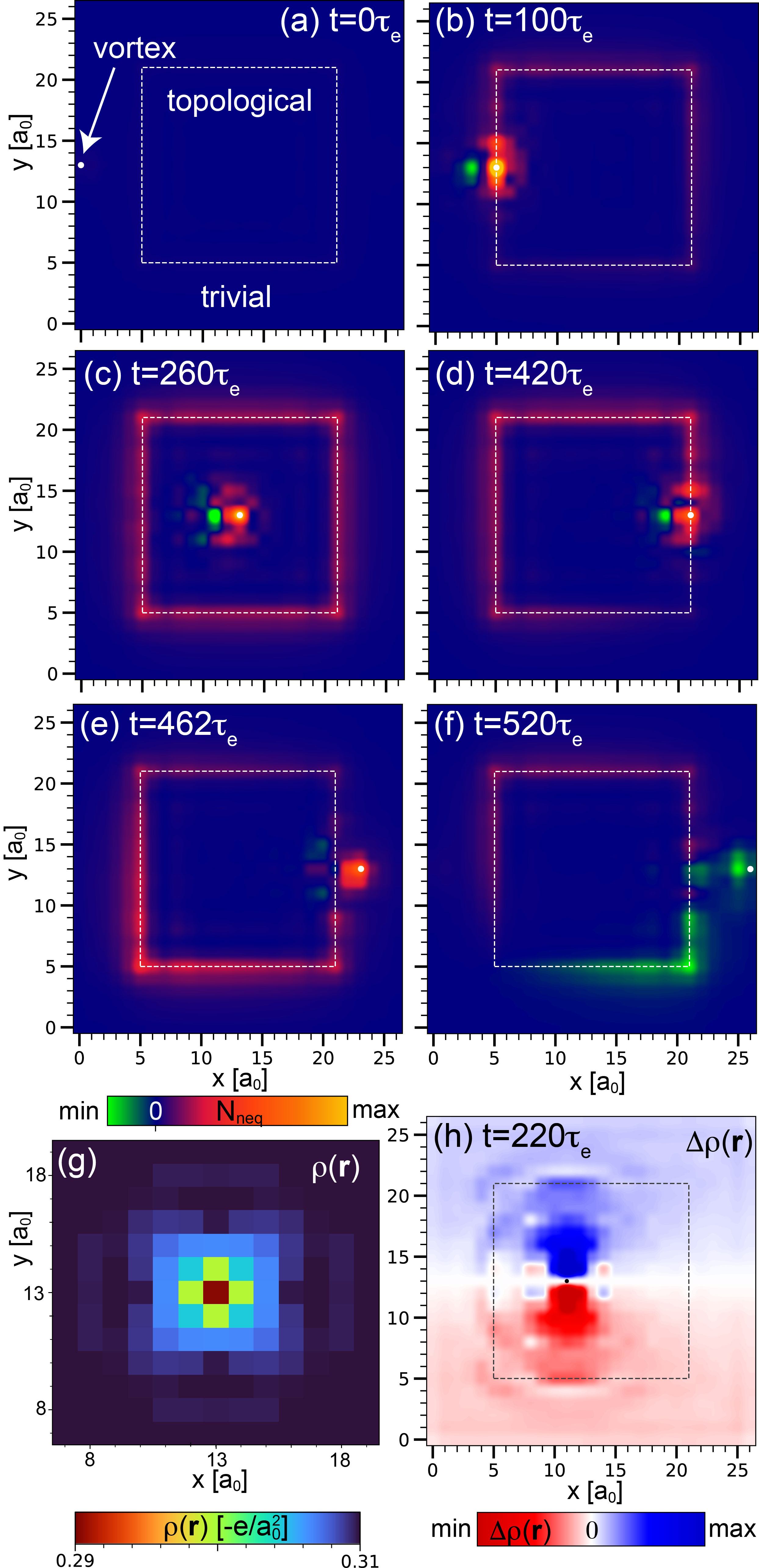}
    \caption{(a)-(f) $N_\mathrm{neq}$ for various times during the motion of the vortex from the trivial into the topological domain. (g) Spatial charge density around a single stationary vortex core. (h) Spatial charge density difference reflecting the Hall effect associated with the motion of the charge surrounding the vortex. Parameters for the topological and trivial domain are $(\mu, J) =(-3.8,2.0)t_e$ and $(\mu, J) =(-3.8,0.6)t_e$, with $(\alpha, \Delta) = (0.8,1.2)t_e$ and $t_V=20\tau_e$. This value of $\Delta$ was used here to ensure a sufficiently small coherence length.}
    \label{fig:Fig4}
\end{figure}

The above results raise the intriguing possibility to study how an MZM emerges in the core of a vortex, when the vortex is moved through a domain wall from a trivial into a topological domain. To investigate this question, we utilize a recently developed non-equilibrium formalism \cite{Bedow2022,Buss2025} to compute the (non-equilibrium) spatially, energy- and time-dependent density of states, $N_\mathrm{neq}({\bf r},\omega,t)$ which  visualizes the vortex motion. We previously showed \cite{Bedow2022} that $N_\mathrm{neq}$ is proportional to the time-dependent differential conductance, $\mathrm{d}I(V,t)/\mathrm{d}V$, measured in STM experiments (for details, see Appendix D). Since the self-consistent calculation of the time-dependent magnitude and phase of the SCOP in the presence of a moving vortex is computationally too demanding, we assume a rigid spatial profile for  $\Delta_{\bf r}(t)$ and $\phi({\bf r},t)$ that moves with the vortex core, as discussed in Appendix D. Below all times are given in units of $\tau_e = \hbar/t_e$ such that for typical values of $t_e$ of a few hundred meV, $\tau_e$ is of the order of a few femtoseconds. The time for the vortex motion from one site to the next is given by $t_V$ (for details, see Appendix C).

In Fig.~\ref{fig:Fig4}(a)-(e), we plot the zero-energy $N_\mathrm{neq}$ for various times during a process in which a vortex is moved from the trivial to the topological domain, the latter being in the form of an island, indicated by a dashed white line (the full time-dependence of $N_\mathrm{neq}$ is shown in Supplementary Movie 1). At the initial time $t=0$, the vortex is located in the trivial domain. The topological island possesses a Majorana edge mode, with the lowest pair of MEMs located at energies $\pm \epsilon_0 \not = 0$ due to the finite size of the island (note that $\epsilon_0 \rightarrow 0$ with increasing size of the island \cite{Rachel2017}). As the vortex moves through the domain wall [see Fig.~\ref{fig:Fig4}(b)], one of the low-energy MEMs is transferred as a localized MZM to the vortex, while its partner remains at the domain wall. When the vortex is moved further into the topological region, and the separation between the MZM and its partner MEM is increased, the reduced hybridization leads to a decrease in their energies $\epsilon_0 \rightarrow 0$, visible as a substantial increase in the spectral weight in the zero-energy $N_\mathrm{neq}$ [see Fig.~\ref{fig:Fig4}(c)].  
When the vortex is moved out of the topological domain on the opposite domain wall, the process is reversed and the MZM is transferred back into the domain wall as a MEM [see Fig.~\ref{fig:Fig4}(d),(e)]. This transfer is completed only when the vortex has exited the topological domain and moved sufficiently far away from the domain wall [see Fig.~\ref{fig:Fig4}(f)]. Some interesting effects occur due to the non-zero velocity with which the vortex moves. In particular, we find that when the vortex enters [see Fig.~\ref{fig:Fig4}(b)] or exits [see Fig.~\ref{fig:Fig4}(d)-(f)] the topological domain, the spectral weight of the low-energy MEM is not isotropic, and does not possess a mirror-symmetry with regards to the $x$-axis. Indeed, the full time dependence of $N_\mathrm{neq}$ reveals that the MEM spectral weight propagates clock-wise around the domain wall upon vortex entry and exit. This non-equilibrium effect is related to the charge associated with the vortex. Indeed, in Fig.~\ref{fig:Fig4}(g) we plot the charge density around a vortex, which reveals that a non-zero charge is localized around the vortex. Thus, as the vortex is moved through the system, a charge is moving with it in an applied magnetic field, giving rise to the Hall effect \cite{Vargunin2019}. As a result, the mirror symmetry along the axis of motion is broken, allowing for the circular motion of the MEMs spectral weight. In addition, we find that as expected from the Hall effect, a finite charge surplus (depletion) accumulates at the upper (lower) boundaries of the system, respectively, which are parallel to the direction of the charge motion, as shown in Fig.~\ref{fig:Fig4}(h). \\

{\bf Discussion.~} We demonstrated that a heterogeneous structure of topological and trivial domains can provide an explanation for the experimental observation that not all vortices in \FSTx host Majorana zero modes. We predicted that an STM experiment conducting a $\mathrm{d}I/\mathrm{d}V$ line cut between a topological and trivial vortex should necessarily observe a domain wall hosting a MEM between them. In contrast, $\mathrm{d}I/\mathrm{d}V$ line cuts between vortices located in the same type of domain -- topological or trivial -- even when possessing different chemical potentials or strength of ferromagnetic moments, should not observe an MEM.  Finally, we showed that by moving a vortex in real time from a trivial to a topological domain, an MEM, delocalized along the domain wall, is transferred to the vortex as a localized MZM. We note that our scenario also provides an explanation for the experimentally observed increase in the number of trivial vortices with increasing magnetic field \cite{Machida2019}. Indeed, we previous predicted that while for weaker ferromagnetism, the system can be in a topological phase with an odd Chern number, which harbors vortex core MZMs (see Supplementary Material in Ref.~\cite{Mascot2022}), with increasing strength of the ferromagnetic ordering, the system can move either into a trivial phase, or a phase with an even Chern number, both of which do not possess a zero-energy mode in the vortex core.

Finally, we note that the above discussed domain structure is qualitatively different from the one considered in Ref.~\cite{Wu2021} to explain the presence and absence of vortex core MZMs; while in our case the vortices with and without an MZM are located in a strong topological and trivial superconducting domain, respectively, both types of domains considered in Ref.~\cite{Wu2021} are topological. \\

{\bf Acknowledgements. ~}
We would like to acknowledge stimulating discussions with M. Buss and J. Hoffman. This work was supported by the U.\ S.\ Department of Energy, Office of Science, Basic Energy Sciences, under Award No.\ DE-FG02-05ER46225.

\appendix

\section{A. Self-consistent Calculation of the Superconducting Order Parameter}

The Hamiltonian in the absence of a magnetic field with periodic boundary conditions is given by
\begin{equation}
    \mathcal{H} = \sum_{ {\bf r}, {\bf r}', \sigma, \sigma'} h(\mathbf{r}, \mathbf{r}', \sigma , \sigma') c^{\dagger}_{\mathbf{r} \sigma} c^{}_{\mathbf{r}' \sigma}
    + \frac{V}{2} \sum_{ {\bf r}, {\bf r}', \sigma, \sigma'} c^{\dagger}_{{\bf r} \sigma} c^{\dagger}_{{\bf r}' \sigma'} c^{}_{{\bf r}' \sigma'} c^{}_{{\bf r} \sigma} \; .
\end{equation}
In the mean-field approximation assuming $s$-wave superconductivity, this Hamiltonian becomes
\begin{equation}
\begin{aligned}
    \mathcal{H} =& \; \sum_{{\bf r}, {\bf r}', \sigma, \sigma'} h(\mathbf{r}, \mathbf{r}', \sigma , \sigma') c^{\dagger}_{ \mathbf{r} \sigma} c^{}_{\mathbf{r}' \sigma} \\
    &+ \frac{1}{2} \sum_{\mathbf{r},\mathbf{r}'} \left( \Delta (\mathbf{r}) c^{\dagger}_{{\bf r} \uparrow}  c^{\dagger}_{{\bf r} \downarrow} + \text{h.c.}  \right) + E_\mathrm{MF} \; ,
\end{aligned}
\end{equation}
where the superconducting order parameter $\Delta(\mathbf{r})  = V \langle c_{\mathbf{r} \downarrow} c_{\mathbf{r} \uparrow} \rangle$ is computed self-consistently. Using the Bogoliubov transformation, this expectation value can be evaluated as 
\begin{equation}
\begin{aligned}
    \Delta(\mathbf{r})  =& \; V \sum_n \left( f(E_n) V^{*}_{n,\downarrow}(\mathbf{r}) U_{n,\uparrow}(\mathbf{r}) \right. \\
    &\left. + f(-E_n) U_{n,\downarrow}(\mathbf{r}) V^{*}_{n,\uparrow}(\mathbf{r}) \right)
\end{aligned}
\end{equation}
with the Fermi distribution function $f(E)$. Here, the coherence factors $U_{n,\sigma}(\mathbf{r}), V_{n,\sigma}(\mathbf{r})$ are determined from diagonalizing the real-space Hamiltonian in Eq.~\ref{eq:ham} using the spinor
\begin{equation}
    \psi_{\mathbf{r}}^\dagger = (c^{\dagger}_{\mathbf{r}, \uparrow}, c^{\dagger}_{\mathbf{r}, \downarrow}, c_{\mathbf{r}, \downarrow}, -c_{\mathbf{r}, \uparrow}) \; ,
\end{equation}
which gives us the eigenvalues $E_n$ and eigenvectors $\psi_n(\mathbf{r}) = (U_{n,\uparrow}(\mathbf{r}), U_{n,\downarrow}(\mathbf{r}), V_{n,\downarrow}(\mathbf{r}), V_{n,\uparrow}(\mathbf{r}))$.
We implement the domain walls so that they have a width $W=4a_0$.

\section{B. Peierls and Gauge Phases due to the Presence of a Magnetic Field}

We incorporate the magnetic field using the Peierls phase
\begin{equation}
\begin{aligned}
    \theta(\mathbf{r, r'}) &= \frac{\pi}{\phi_0} \int_{\mathbf{r}}^{\mathbf{r'}} \mathbf{A} \cdot \mathrm{d}\mathbf{l} \; ,
\end{aligned}
\end{equation}
where $\phi_0 = \frac{h}{2e}$ is the superconducting flux quantum. Using the symmetric gauge ${\bf A} = \frac{1}{2} {\bf B} \times {\bf r}$, this can be evaluated as 
\begin{align}
    \theta(\mathbf{r, r'}) & = \frac{\pi}{2 \phi_0}  \mathbf{B} \cdot \left(\mathbf{r} \times \mathbf{r}' \right) \; .
\end{align}

The Peierls phase modifies the hopping amplitude and Rashba SOC according to (see Eq.~\ref{eq:ham} in the main text)
\begin{subequations}
\begin{align}
    t_e &\rightarrow t_e e^{\mathrm{i}\theta({\bf r}, {\bf r'})}\; ,  \\
    \alpha &\rightarrow \alpha e^{\mathrm{i}\theta({\bf r}, {\bf r'})}\; .
\end{align}
\end{subequations}
In the main text, we consider a system with $N_x (N_y)$ sites in the $x$-($y$-)direction (the supercell) and periodic boundary conditions (PBCs). For such a periodic system, we require that the BdG equations be gauge-invariant, implying that they are invariant under a translation by the spatial vector 
\begin{equation}
    {\bf R} = m_1 {\bf u}_1 + m_2 {\bf u}_2 \; ,
\end{equation}
where $m_1$,$m_2$ are integers and ${\bf u}_1 = N_x {\bf a}_1$, ${\bf u}_2 = N_y {\bf a}_2$. The vector potential in the symmetric gauge changes under a translation by ${\bf R}$ as
\begin{equation}
    {\bf A}({\bf r} + {\bf R}) = {\bf A}({\bf r}) + {\bf A}({\bf R}) \; ,
\end{equation}
which we can interpret as the gauge transformation
\begin{equation}
    {\bf A}({\bf r} + {\bf R}) = {\bf A}({\bf r}) + \nabla \chi({\bf r}, {\bf R}) \; ,
\end{equation}
with 
\begin{equation}
\begin{aligned}
    \chi({\bf r}, {\bf R}) &= {\bf A}({\bf R}) \cdot {\bf r} + C({\bf R})
    = \frac{1}{2} ({\bf H} \times {\bf R}) \cdot {\bf r} + C({\bf R}) \;.
    \label{eq:gauge_one_vortex}
\end{aligned}
\end{equation}
Here, $\bf C(R)$ is a function of $ \mathbf{R} = m_1 \mathbf{u}_1 + m_2 \mathbf{u}_2$, which can be evaluated as
\begin{align}
    C(\mathbf{R}) &= C(m_1 \mathbf{u_1} + m_2 \mathbf{u_2}) \nonumber \\
    &= m_1 C(\mathbf{u}_1) + m_2 C( \mathbf{u}_2 ) - \frac{1}{2} m_1 m_2 \ |\mathbf{u}_1 \times \mathbf{u}_2| \hat{z} \cdot \mathbf{B} \nonumber \\
    &= m_1 C(\mathbf{u}_1) + m_2 C( \mathbf{u}_2 ) - \frac{1}{2} m_1 m_2 \ \mathbf{S} \cdot \mathbf{B} \nonumber \\
    &= m_1 C(\mathbf{u}_1) + m_2 C( \mathbf{u}_2 ) - \frac{1}{2} m_1 m_2 n_{\phi} \phi_0
\end{align}
where $\displaystyle \mathbf{B} = (0,0,n_{\phi} \phi_0 \:/\: S) $ is a constant magnetic field in the $z$-direction with $n_{\phi}$ flux quanta per unit cell of area $S$. 
The Peierls phase for a translation by $\bf R$ is then given by

\begin{align}
    \theta (\mathbf{r + R \ r' + R}) &= \theta (\mathbf{r,r'}) + \phi(\mathbf{r',R}) - \phi(\mathbf{r,R}) \; ,
\end{align}
where $\phi({\bf r},{\bf R}) = \frac{\pi}{\phi_0} \chi({\bf r},{\bf R})$.

We can determine the coefficients $C(\mathbf{u}_1)$ and $C( \mathbf{u}_2)$ by considering the effect of a translation by ${\bf R}$ on the superconducting order parameter. In particular, the BdG equations yield for the superconducting order parameter 

\begin{equation}
    \Delta ({\bf r}+{\bf R}) = \Delta({\bf r}) e^{ -2i \phi({\bf r},{\bf R}) } \; .
\end{equation}
By imposing that the SCOP be continuous across the boundaries of a supercell into the next supercell, we can then find suitable values for $C(\mathbf{u}_1)$ and $C( \mathbf{u}_2)$.
For hopping processes outside of the supercell, the Peierls and gauge phase together then modify the hopping amplitude as 
\begin{equation}
    t_e \to t_e \exp\left(\mathrm{i}\theta({\bf r},{\bf r}^\prime) \right) \exp\left(\mathrm{i} \phi({\bf r}^\prime, -{\bf R})\right) \; .
\end{equation}
and the Rashba spin-orbit coupling as
\begin{equation}
    \alpha \to \alpha \exp\left(\mathrm{i}\theta({\bf r},{\bf r}^\prime) \right) \exp\left(\mathrm{i}\phi({\bf r}^\prime, -{\bf R})\right) \; .
\end{equation}

\section{C. Time-dependent Vortex Motion}

To describe the movement of a vortex from one lattice site to the next in a time $t_V$ (for horizontal, vertical, and diagonal directions), we utilize a smoothed-out Heaviside-function given by
\begin{equation}
    s(t, t_0, t_V) = 
    \begin{cases} 
    0 ,& t < t_0 \\
    \sin^2\left( \frac{t-t_0}{t_V}\right) ,& t_0 \leq t \leq t_0 + t_V \\
    1 ,&  t > t_0 + t_V
    \end{cases} \; .
\end{equation}
To move the vortex core, for example, along the $x$-direction from a site ${\bf R}(t_0)$ at time $t_0$ to ${\bf R}(t_0)+{\hat {\bf x}}a_0$ at time $t_0+t_V$, we change its position as 
\begin{equation}
    \mathbf{R}(t) = \mathbf{R}(t_0) + {\hat {\bf x}}a_0 s(t,t_0, t_V)
\end{equation}
and analogously for movements along the $y$- or diagonal directions.\\

The motion of the magnetic vortex leads to a time dependence in the magnitude $|\Delta_{\bf r}(t)|$ and phase $\phi({\bf r}, t)$ of the superconducting order parameter. The former is described by 
\begin{equation}
    |\Delta_{\bf r} (t)| = \Delta_0 \sin^2\left(\frac{\pi}{2 R_V} \min\left[ \left|{\bf r} - {\bf R}(t)\right|, R_V\right]\right) \; ,
\end{equation}
whose spatial profile describes the reduction of the superconducting order parameter within a radius $R_V$ of the time-dependent location of the vortex core at site ${\bf R}(t)$, with $|\Delta_{\bf r} (t)| =0 $ at the center of the vortex. Moreover, the smoothed-out Heaviside function guarantees that the reduction of the superconducting order parameter's magnitude to zero at the vortex core occurs smoothly, which is necessary for the adiabatic evolution of the vortex motion. While for this spatial form, $|\Delta_{\bf r}(t)|$ decreases less steeply around the vortex center than that obtained in self-consistent calculations \cite{Smith2016}, it was previously shown that the occurring vortex core MZMs are robust with respect to the spatial profile of $|\Delta_{\bf r}(t)|$ \cite{Nagai2014}. Thus, while the spatial form of $|\Delta_{\bf r}(t)|$ utilized here leads to a larger MZM localization length in comparison to that obtained self-consistently, this does not affect the results presented in Fig.~\ref{fig:Fig4} of the main text. 

To describe the time dependence of $\phi({\bf r}, t)$, we first determine its value when the vortex is centered at a lattice site or at the center of plaquette of four lattice sites, at site $t_0$, given by 
\begin{align}
    \phi({\bf r}, t_0) &= - \sum_{p,q=-\infty}^{\infty} \arctan\left( \frac{r_y - R_{y}(t_0) - p N_y}{r_x - R_{x}(t_0) - q N_x} \right) \; ,
    \label{eq:SCOP_phase}
\end{align}
where ${\bf r}=(r_x,r_y)$ and ${\bf R}=(R_{x},R_{y})$. This form of the superconducting order parameter approximate well that obtained from the self-consistent calculations \cite{Vafek2001,Nagai2012,Smith2016} and ensures continuity of the phase across the boundary of the magnetic supercell. $N_x$ and $N_y$ are the numbers of sites in the supercell in the $x$- and $y$-directions, and $p,q$ run over all copies of the supercell due to the periodic boundary conditions. We then interpolate $\phi({\bf r}, t)$ between times $t_0$ and $t_0 + t_V$ using the smoothed-out Heaviside function 
\begin{equation}
    \phi({\bf r}, t) = \phi({\bf r}, t_0) + \left[\phi({\bf r}, t_0+t_V)-\phi({\bf r}, t_0) \right]
    s(t,t_0, t_V) \; .
\end{equation}
In order to make the time-dependent computations feasible, we utilize a larger value of $\Delta$ in Fig.~\ref{fig:Fig4}, such that the superconducting coherence length $\xi \sim \Delta / \epsilon_F$ is shorter, which allows us to consider smaller system sizes.

\section{D. Time-Dependent Physical Quantities}
\label{sec:timedep_observables}

In order to visualize the vortices in real time and space, we employ the energy-, time- and spatially-resolved non-equilibrium density of states $N_\mathrm{neq}({\bf r}, \sigma, t, \omega) = -\frac{1}{\pi} {\rm Im} \, G^\mathrm{r}({\bf r},{\bf r}, \sigma, t, \omega)$, 
which we previously showed to be proportional \cite{Bedow2022} to the time-dependent differential conductance $\mathrm{d}I(V,{\bf r},t)/\mathrm{d}V$ measured in STM experiments. 
Therein, the time- and frequency-dependent retarded Greens function $\hat{G}^\mathrm{r}$ is calculated from the differential equation
\begin{equation}
   \left[ \mathrm{i} \frac{\mathrm{d}}{\mathrm{d}t} +  \omega  + \mathrm{i} \Gamma - {\hat H}(t)\right] {\hat G}^\mathrm{r} \left(t, \omega \right) = {\hat 1} \; .
\end{equation}

We compute the time-dependent electronic charge density of the ground state $\ket{\psi(t)}$ \cite{Hodge2025_2,Bedow2025} using
\begin{equation}
\begin{aligned}
    \rho({\bf r}, \sigma, t) &= -e \braket{\psi(t)|c^\dagger_{{\bf r}, \sigma} c_{{\bf r}, \sigma}|\psi(t)} \\
    &= -e \left(\widehat{\tilde{V}}(t) \widehat{\tilde{V}}^\dagger(t) \right)_{({\bf r}, \sigma),({\bf r}, \sigma)} \; .
\end{aligned}
\end{equation}

\end{document}